**In-situ electron backscatter diffraction of thermal cycling in a single grain Cu/Sn-3Ag-0.5Cu/Cu solder joint**

Tianhong Gu*, Yilun Xu, Christopher M. Gourlay, and T. Ben Britton

Department of Materials, Imperial College London, SW7 2AZ. UK

*Corresponding author: t.gu15@imperial.ac.uk

The heterogeneous evolution of microstructure in a single Cu/SAC305/Cu solder joint is investigated using in-situ thermal cycling combined with electron backscatter diffraction (EBSD). Local deformation due to thermal expansion mismatch results in heterogeneous lattice rotation within the joint, localised towards the corners. This deformation is induced by the constraint and the coefficient of thermal expansion (CTE) mismatch between the β-Sn, $Cu_6Sn_5$ and Cu at interfaces. The formation of subgrains with continuous increase in misorientation is revealed during deformation, implying the accumulation of plastic slip at the strain-localised regions and the activation of slip systems $(110)[\bar{1}11]/2$ and $(1\bar{1}0)[111]/2$.

**Keywords:** Pb-free solder, Polygonization, In-situ thermal cycling, and Microstructure evolution.

The failure of electronics in service is often due to thermomechanical failure of the solder joints. This drives our need to understand the mechanisms leading to thermomechanical fatigue failure, to be able to predict the joint microstructure that is most resistant to thermal cycling and to develop ways of generating the optimum microstructure through alloy design and processing.

SAC305 is a Sn-based alloy containing ~96.5% β-Sn phase [1]. β-Sn has a body-centred tetragonal (BCT) structure and behaves anisotropically in thermophysical properties. For example, the coefficient of thermal expansion (CTE) is double along the c-direction than along a = b [2] and also stiffness is triple along the c-direction than a = b [3]. Furthermore, in a solder joint the SAC305 ball is attached through intermetallic and metallisation layers to a polymer based printed circuit board (PCB) and, often silicon, which have significant differences in CTE to the solder ball. Thus, solder joints are particularly sensitive to thermal cycling.

The major failure mode of solder joints in thermal cycling is cracking during heterogeneous microstructure evolution, i.e. polygonization and crack nucleation, in the bulk solder near the solder/substrate interface on the Si chip-side, observed in ref [4-19]. The crack propagates in the region where the mismatch of CTE is large and the induced thermal strain is high in comparison to the surroundings (larger misfit on the Si chip side than PCB side of joint [2, 20-22]). This CTE mismatch between the major constituents (solder / IMC layers / copper / PCB / silicon) leads to localised plastic deformation, this results in high local dislocation density and the stored energy may be released by recovery, recrystallization and grain growth in the β-Sn lattice. Ultimately this can provide an energetically favourable site for cracking within the β-Sn [5, 9, 14-16, 18, 23].

Due to the low melting temperature of tin based solders, the recovery process takes place during deformation, where the stored energy or the driving force can be released by forming subgrains with misorientation between 2 - 15° [7, 19, 24]. The tested solder joints deform by continuous lattice rotation, caused by the operation of certain slip systems [25].

The generation of {110} slip traces is often associated with polygonization and recrystallization [6-8, 24]. Zhou et al. [7] analysed slip traces from SAC305 solder joints with [110] orientation close to the loading direction and demonstrated surface roughness enhanced at the strain-localised region with formation of subgrains, which are attributed to the generation of dislocations and the operation of slip system: $(1\bar{1}0)[001]$ with gradual lattice rotation around the [110] axis. With increasing straining, recrystallized grains develop progressively with greater misorientation [7]. Han et al. [24] carried out thermal shock testing for a multi-crystal SAC305 solder joint and found that the β-Sn deforms by subgrain rotation with [001] and [100] rotational axes for the chip and PCB side respectively. Bieler et al. [6] showed with SAC305 solder joints that the {110}[001] slip systems are the most easily activated when the β-Sn c-axis is aligned with the loading direction and stimulates recrystallization at the strain localised region. Furthermore, Telang et al. [8] reported continuous recrystallization with the formation of {110} slip plane traces in a Sn-Ag joint during thermal cycling. Based on analysis using EBSD and 3D finite element modelling (FEM), newly recrystallized orientations grew to consume initial orientation(s) and the rate of recrystallization was related to the elastic strain energy driving force [8].

While significant advances have been made in understanding microstructure and damage development during thermal cycling, a range of questions remain about the early stages of microstructure evolution during thermal cycling, e.g. When and where does the microstructural deformation start? How does heterogeneous deformation arise? Where does the strain accumulate? How do the β-Sn grains (re)orientate? If we can characterise and understand these deformation processes, then we may be able to predict the optimum microstructures for reliable electronic components.

This paper explores these questions using an in-situ heating stage together with EBSD during the early stages of thermal cycling. We study ball grid array (BGA) scale Cu/SAC305/Cu solder joints containing a single β-Sn grain, and focus on the localisation of plastic deformation and microstructure evolution with increasing number of thermal cycles.

A sandwich geometry solder joint (Cu/SAC305/Cu) was fabricated for in-situ thermal cycling with SEM analysis. The sample dimensions were $8 \times 8 \times 1.5$ mm$^3$ consisting of a $5 \times 5$ array of solder balls with ball pitch size 1 mm and pad size 0.5 mm. The solder balls were ~500 μm in diameter and connected with two Cu substrates (~500 μm in thickness, which constrains the freedom of solder ball during thermal cycling and gives a loading direction parallel to the plane of substrate). The solder balls were made of SAC305 solder ingots with composition Sn-3.052Ag-0.504Cu in wt% and main impurities of 0.029Pb, 0.007Sb, <0.003Fe, <0.003Bi, and Zn, Cd, As, Al all <0.001 wt% as measured by XRF spectroscopy. Ingot was rolled into foil, punched into thin disks and then melted on a hot plate with a mildly activated rosin flux (RM-5) to form near-spheres due to surface tension [26]. After that, the solder balls were soldered to the Cu substrates with the flux and then reflowed through a conveyor oven (LFR400HTX TORNADO Surface Mount Technology, UK) [26]. All samples were cleaned in an ultrasonic bath with ethanol to remove flux residues and were then submerged in acetic acid for 20 minutes to remove the solder mask layer.

The solder joint array was polished without mounting using a Gatan Disk Grinder 623. The microstructure of the joint was revealed by standard mechanical polishing to a 0.05 μm colloidal silica (OPS) finish. The quality of the cross-sectional surface of the solder joint was then improved by ion-beam milling (PECSII) at 1 and 0.5 keV with a tilt angle of 3 and 1° respectively for a high spatial resolution EBSD scan.

In-situ thermal cycling together with EBSD scanning was carried out with the Gatan Murano 525 heating stage inside a Quanta FEI 450 SEM. The solder joint was thermally cycled from 20 to 150 °C with a dwell time of 10 minutes at the temperature extremes. The heating rate was ~12.5 °C/min. and the cooling rate was ~4.5 °C /min (the profile is chosen according to the IPC9701 [27]). An EBSD map was scanned of the whole joint at room temperature after every 12 thermal cycles with a step size of 1.8 µm and the solder / substrate interface was scanned with a step size of 0.2 µm after 36 cycles. Between every 12 cycles, the joint was detached from the hot plate and put inside a vacuum desiccator to avoid oxidation. The EBSD data shown was analysed using Bruker ESPRIT 2.1 with post-processed "absorb surrounded zero solutions".

*Figure 1* shows a single β-Sn grain solder joint with its c-axis almost within the plane of the substrate (approximately 20°), which gives the largest in-plane CTE mismatch with the Cu substrate (*Table 1*). We note that the non-linear evolution of CTEs may further enhance this mismatch [28].

As the joint is thermally cycled, a gradual change in colour of the IPF-LD maps at the corners of the joint from green to cyan after 36 (*Figure 1*b) and 60 cycles (*Figure 1*c) occurs compared to the maps at 12 cycles (*Figure 1*a), indicating the orientation is rotated gradually as the number of cycles increases. This is caused by the CTE mismatch between the three major phases (the CTE values for β-Sn, $Cu_6Sn_5$ and Cu are given in *Table 1*), which generates largely thermal strain and causes lattice rotation at the interface in the solder matrix (due to the softness of β-Sn). After 84 thermal cycles, the accumulated plastic deformation causes noticeable reduction in indexing at the corners of the joint with no EBSD pattern detected (*Figure 1*d).

The in-grain misorientation maps in *Figure 2* show two intersecting shear bands that extend from the corners of the joint to form a cross. The highest misorientation appears at the corners (MO av. colour key 0 – 15°). It is observed that a slight lattice rotation occurs from the centre to the corner of the joint. Along each band, the minimum misorientation is at the centre of the joint and increases in magnitude as moving from the centre towards the joint corners, along each diagonal deformation band (*Figure 2*a – 2d), i.e. from the centre (location 1) to the corner of the joint (location 2 and 3), the orientation rotates counter-clockwise increasingly, while the orientation rotates clockwise as moving down the cross band (location 4 and 5) as illustrated in *Figure 2*f. After 84 cycles (*Figure 2*d), the non-indexed positions appear at the corners, where relatively higher misorientations are observed than at previous cycles (*Figure 2*a – 2c).

Generally, the misorientation increases with increasing thermal cycling shown as an increase in peaks to larger angles in the misorientation curves at higher number of cycles (*Figure 2*e). The misorientation angle ($\Delta\theta$) vs. position (x) curves (*Figure 2*e) are drawn according to the red dashed line in each MO av. map). They show the rotation angle increases as the distance becomes farther from the centre of the joint and the curve has larger gradient at higher thermal cycle number. *Figure 2*f plots the unit cell orientations with increasing number of cycles from the measured Euler angles. It can be seen that the β-Sn rotates around the [001] rotational axis and the direction of rotation is indicated with blue arrows.

When the solder is deformed by thermal cycling, the lattice misorientation within the solder joint is increased continuously by lattice rotation resulting from crystal slip along a single rotational axis (*Figure 2*f). This lattice rotation indicates higher straining and more highly stored energy in the region, which causes formation of significant subgrain boundaries and is vital for generation of HAGBs (recrystallization) and crack nucleation in the later stages of thermal cycling [7, 8, 14-16, 18].

The microstructure of the β-Sn / $Cu_6Sn_5$ / Cu reaction zone is shown in the *Figure 3* after 12 thermal cycles. The four phases: β-Sn (red), $Cu_6Sn_5$ (blue), $Ag_3Sn$ (green) and Cu (yellow) are identified clearly in the phase map (*Figure 3*a). Only a few $Ag_3Sn$ particles are indexed in the region because of its fine size (sub-micrometre scale). A layer of $Cu_6Sn_5$ and $Cu_6Sn_5$ in the bulk solder can be seen in *Figure 3*, which are approximately 1 – 5 µm in size. Near the $Cu_6Sn_5$ layer, the high misorientation is concentrated in the β-Sn phase (*Figure 3*c – 3e) with significant formation of subgrains (*Figure 3*b).

The deformation of the solder joint is induced at the corner of the solder matrix during thermal cycling at the β-Sn, $Cu_6Sn_5$ and Cu interface. This is caused by the CTE mismatch between the 3 phases (*Table 1*), which systematically varies the stress and strain build-up during thermal loading near the interface of the solder joint and results in microstructural evolution in the solder matrix.

The lattice rotation during deformation can be evaluated based on the rotation matrix from the initial crystallographic orientation to the deformed one, which gives:

$$\boldsymbol{R} = (\boldsymbol{R}_{ini})^{-1}\boldsymbol{R}_{rot} \quad (1)$$

Where $\boldsymbol{R}_{ini}$ and $\boldsymbol{R}_{rot}$ are the initial and deformed rotation matrices from the reference configuration that are calculated from the EBSD measured Euler angles. The rotation axis $\boldsymbol{u}_{exp}$ of the lattice rotaton matrix $\boldsymbol{R}$ is then given by:

$$\boldsymbol{u}_{exp} = \begin{pmatrix} u_x \\ u_y \\ u_z \end{pmatrix} = \begin{pmatrix} \boldsymbol{R}(3,2) - \boldsymbol{R}(2,3) \\ \boldsymbol{R}(1,3) - \boldsymbol{R}(3,1) \\ \boldsymbol{R}(2,1) - \boldsymbol{R}(1,2) \end{pmatrix} \quad (2)$$

The analytical rotation axis $\boldsymbol{u}^{(\alpha)}_{analyt}$ is determined by the $\alpha^{th}$ slip system with slip normal $\boldsymbol{n}^{(\alpha)}$ and slip direction $\boldsymbol{s}^{(\alpha)}$ of a Sn crystal as:

$$\boldsymbol{u}^{(\alpha)}_{analyt} = \boldsymbol{n}^{(\alpha)} \times \boldsymbol{s}^{(\alpha)} \qquad (3)$$

Hence, the deviation angle $\Delta\omega$ between the experimental measured rotation axis $\boldsymbol{u}_{exp}$ and the rotation axis by a certain slip system $\boldsymbol{u}_{analyt}$ enables to determine which specific slip systems are potentially activated during deformation. The slip system(s) with minimum $\Delta\omega$ is likely to be activated.

The experimental rotation axis is obtained from the location 1 (the centre of the solder given in *Figure 2*a) to the configuration 2 and 4 (shown in *Figure 2*f). The deviation angle $\Delta\omega$ between the experimental and analytical rotational axis is ranked from smallest to largest as plotted in *Figure 4*a and 4c.

The activation of slip systems can be found by analysing $\Delta\omega$ and calculating the weighted Schmid factor ($m_W$) for the crystal orientation, which is measured by EBSD Euler angles. The $m_W$ can be calculated according to the critical resolved shear stress ($\tau_{CRSS}$) and Schmid factor (m), as given in Equation (4).

$$m_W = \frac{m}{\tau_{CRSS}/\tau_{CRSS}^{[001]}} \qquad (4)$$

Where $\tau_{CRSS} = \frac{F}{A}\cos\phi\cos\lambda$ and the Schmid factor, $m = \cos\phi \times \cos\lambda$.

Therefore, by considering both $m_W$ and $\Delta\omega$, slip system #8 (110)[$\bar{1}$11]/2 and #9 (1$\bar{1}$0)[111]/2 are indicated as the most active slip systems for the solder joint under cyclic thermal loading, according to their relatively high $m_W$ and small $\Delta\omega$ among all slip systems (*Table 2*).

Although the slip systems #11 (110)[1$\bar{1}$0], #12 (1$\bar{1}$0)[110] and #15 (100)[011] have a small Δω, they are unlikely to be active due to relatively lower values of $m_W$ and longer slip distances than the slip systems #8 (110)[$\bar{1}$11]/2 and #9 (1$\bar{1}$0)[111]/2, which have been reported by Zhou et al. [29], Kinoshita et al. [30] and Matin et al. [31]. The slip systems presented are based on prior work by Fujiwara et al. [32] and Zhou et al. [29]. The common Sn slip systems [32] with CRSS ratios ($\tau_{CRSS}/\tau_{CRSS}^{[001]}$) [29], m and $m_W$ values are included in Supplementary Table S1.

Moreover, it is found that the activated slip systems are in line with the directions of the two diagonal deformation bands in *Figure 2*, which are also relative to the dendrite microstructure pointing along [110] and [1$\bar{1}$0] in the cross-section of the solder (Figure S2). It is known that [110] and [1$\bar{1}$0] are the dendrite growth directions in β-Sn, where the primary and secondary dendrite arms are indicated with blue line 1 and red line 2 respectively in Figure S2.

In-situ thermal cycling has been performed to investigate the heterogeneous evolution of microstructure in a single β-Sn crystal Cu/SAC305/Cu BGA solder joint. In the early stage of thermal cycling, it has been revealed that heterogeneous deformation was associated with both the constraint by the CTE mismatch between the three major phases (Cu, $Cu_6Sn_5$ and β-Sn) and the anisotropy of β-Sn. The plastic deformation localised at the corners of the β-Sn /$Cu_6Sn_5$ / Cu interface resulted in heterogeneous lattice rotation, i.e. the greater the distance from the centre of the joint the greater the rotation, which is a consequence of the increase in misorientation and formation of subgrains (polygonization) by recovery processes.

The activation of slip systems has been analysed by measuring the lattice rotation angle and calculating the weighted Schmid factor during the microstructural evolution for the crystal orientation of the solder joint. The most likely activated slip system are (110)[$\bar{1}$11]/2 and ($1\bar{1}0$)[111]/2, which produced two diagonal deformation bands in the misorientation maps as illustrated in *Figure 2*, and the evolution in stored energy as a result of further cycling, is likely the precursor to recrystallization and crack propagation.

This work provides new understanding on the early stages of microstructural deformation and the accumulation of stored energy associated with these bands as the precursor to the more widely studied phenomena of recrystallization and crack propagation in thermal cycling.

TG drafted the initial manuscript and conducted the experimental work. YX supported the development of the Matlab code for Sn slip systems. CG and TBB supervised the work equally. TG, CG and TBB contributed to the final manuscript.

Data from this manuscript is available at: <link to be included upon final acceptance>.

TBB would like to thank the Royal Academy of Engineering for his research fellowship. CG would like to thank EPSRC (EP/M002241/11) for funding of his research fellowship. We would like to thank EPSRC (EP/R018863/1) for funding. We would like to thank Nihon Superior for positive support and encouragement to conduct this work. We would like to acknowledge Dr Sergey Belyakov for support in the initial fabrication of the samples and Prof Fionn Dunne for helpful discussions on slip system activity. The microscope and loading frame used to conduct these experiments was supported through funding from Shell Global Solutions and is provided as part of the Harvey Flower EM suite at Imperial.

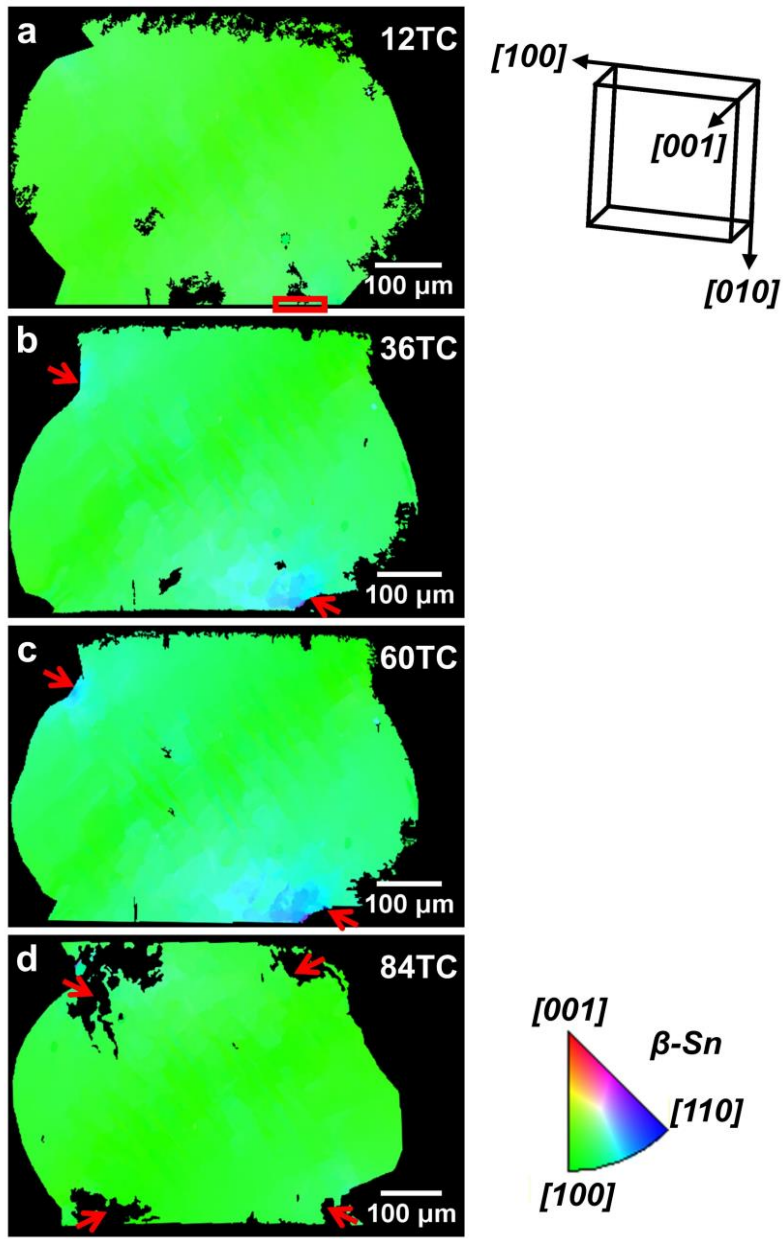

*Figure 1. EBSD IPF-LD (Crystal orientations represented with EBSD inverse pole figure colouring with respect to the loading direction, IPFX) maps showing heterogeneity in microstructural evolution of a single β-Sn grain SAC305 solder joint at different number of thermal cycles, (a) after 12 cycles, (b) after 36 cycles, (c) after 60 cycles and (d) after 84 cycles. (Only the β-Sn has been indexed here).*

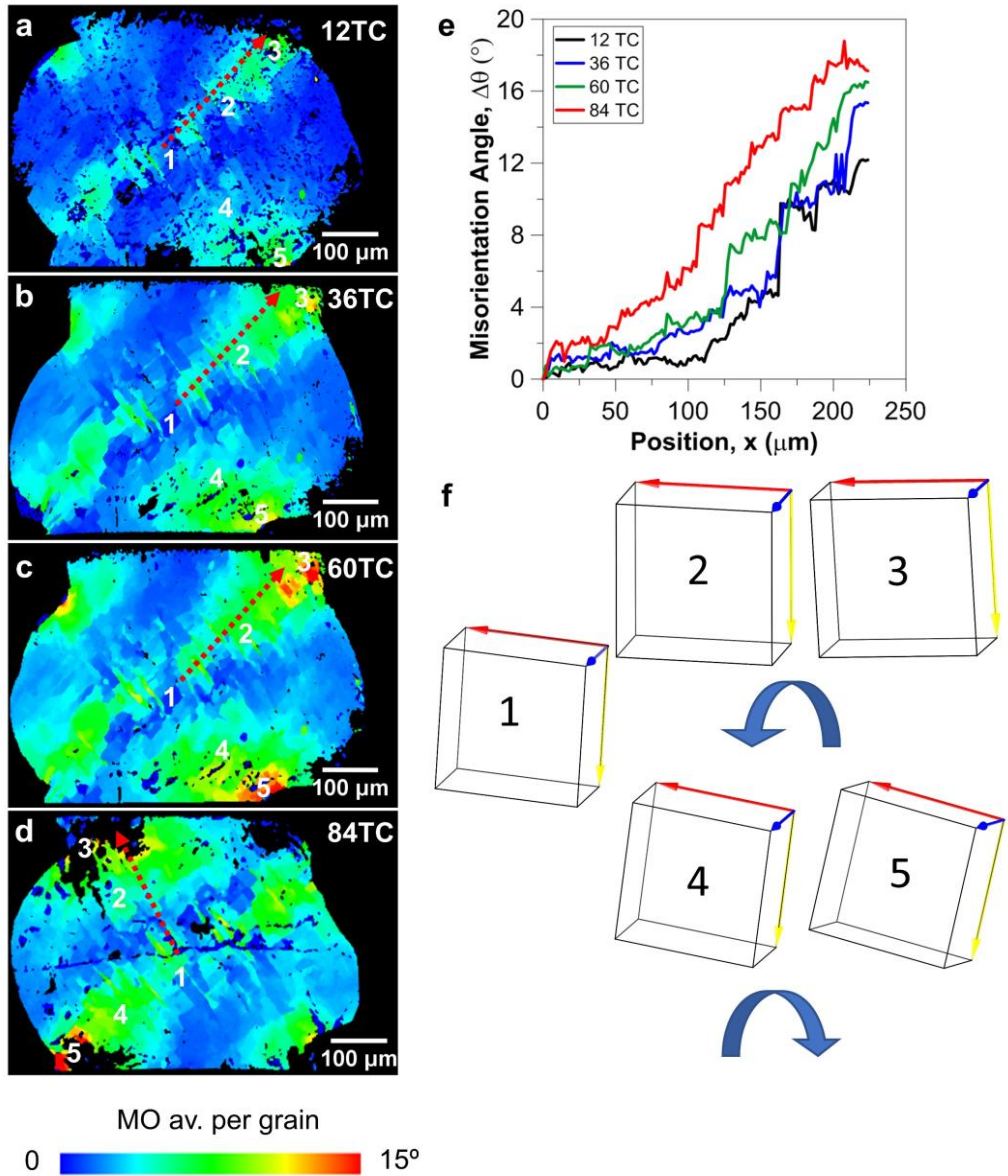

*Figure 2.* EBSD MO av. map showing misorientation evolution of a single β-Sn grain SAC305 solder joint at different number of thermal cycles, (a) after 12 cycles, (b) after 36 cycles, (c) after 60 cycles and (d) after 84 cycles. (e) Orientation rotation angle vs. position curves, the curve is drawn according to the red dashed line on (a-d). (f) Illustrations of the unit cell orientations showing the crystal rotation behaviour, location 1 is at the centre of the solder, location 2 - 5 are the rotated frame of the grain at corresponding locations on the MO maps. (Only the β-Sn has been indexed here).

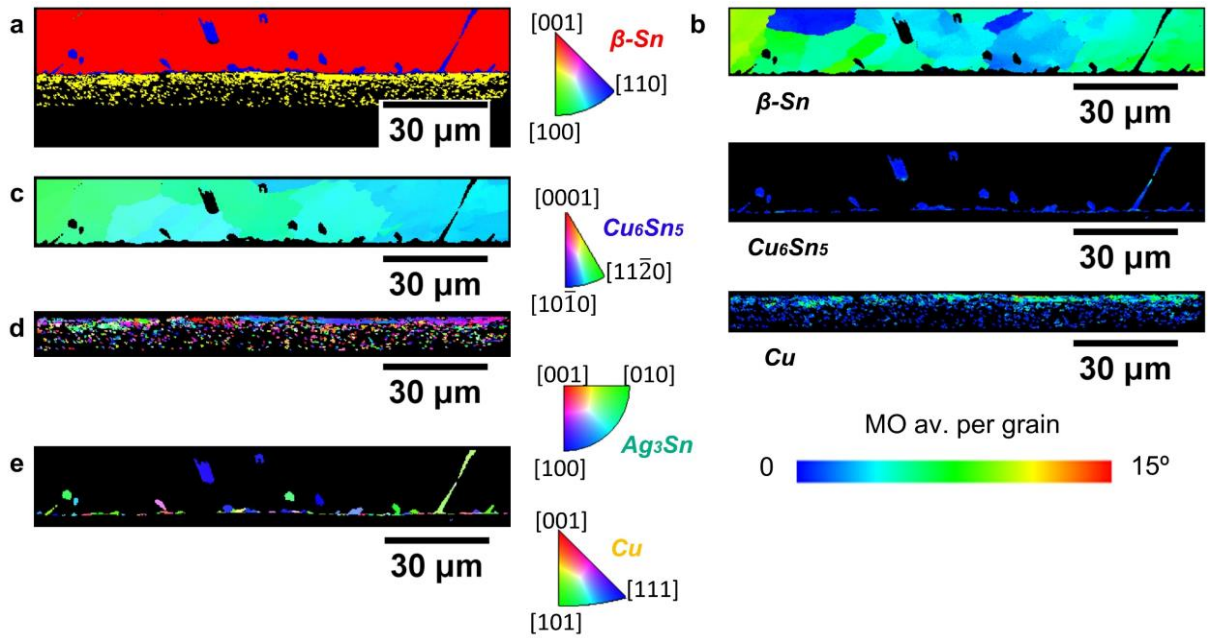

*Figure 3.* EBSD crystal orientation mapping showing the microstructure at β-Sn / Cu$_6$Sn$_5$ / Cu interface after 12 thermal cycles within the red rectangular region in *Figure 1*a. (a) Phase map (red = β-Sn, blue = Cu$_6$Sn$_5$, green = Ag$_3$Sn, yellow=Cu). (b) MO av. map for β-Sn, Cu$_6$Sn$_5$ and Cu respectively. (c-e) IPF-LD maps of β-Sn (c), Cu$_6$Sn$_5$ (d) and Cu (e).

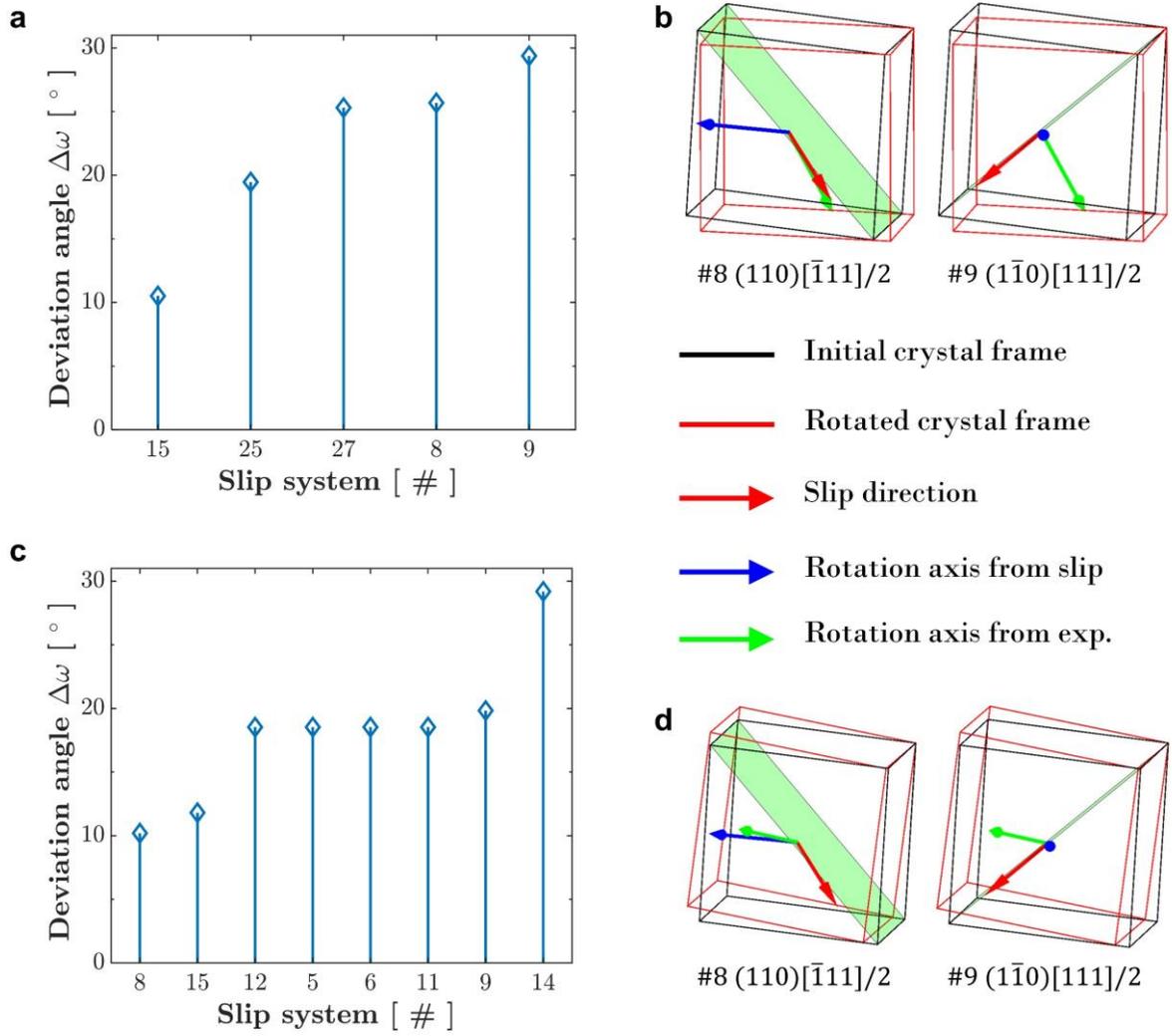

*Figure 4.* Lattice rotation axis analysis of constraint slip for single grain solder joint. (a, c) The distributions of the deviation angle ($\Delta\omega$) between experiental and analytical rotation axes (shortlist to $\Delta\omega \leq 30°$), the EBSD Euler angles are measured from location 1 to 2 (a) and from location 1 to 4 (c) respectively in *Figure 2*f. (b, d) The activated slip systems are illustrated in orientation unit cells with rotational axes, and slip direction labelled, the green shadowed planes represent the slip planes.

| Axis | β-Sn (°C$^{-1}$) [2] | η'-Cu$_6$Sn$_5$ (°C$^{-1}$) [2] | Cu (°C$^{-1}$) [20] |
|---|---|---|---|
| a | 14.195 | 19.195 | 16.5 |
| b | 14.195 | 17.684 | 16.5 |
| c | 28.662 | 18.695 | 16.5 |

*Table 1.* The CTE values of the interface constituents. The values are determined using in-situ powder X-ray diffraction (PXRD) and calculated by fitting the PXRD data into corresponding crystal structure [2, 20, 28]. A typical PCB material is a glass reinforced epoxy FR-4.

| Family | Slip system | | $\Delta\omega_{1-2}$ (°) | $\Delta\omega_{1-4}$ (°) | $m_w$ |
|---|---|---|---|---|---|
| 3 | 5 | (100)[010] | - | 18.5 | 0.046 |
|   | 6 | (010)[100] | - | 18.5 | 0.046 |
|   | 8 | (110)[$\bar{1}$11]/2 | 25.7 | 10.2 | 0.344 |
|   | 9 | (1$\bar{1}$0)[111]/2 | 29.4 | 19.8 | 0.449 |
| 5 | 11 | (110)[1$\bar{1}$0] | - | 18.5 | 0.389 |
|   | 12 | (1$\bar{1}$0)[110] | - | 18.5 | 0.389 |
|   | 14 | (010)[10$\bar{1}$] | - | 29.2 | 0.029 |
|   | 15 | (100)[011] | 10.5 | 11.8 | 0.125 |
| 10 | 25 | (121)[$\bar{1}$01] | 25.3 | - | 0.256 |
|   | 27 | ($\bar{1}$21)[101] | 25.3 | - | 0.093 |

*Table 2.* The shortlisted Sn slip systems [32] in *Figure 4*a and c are given with the deviation angle ($\Delta\omega$) measured from location 1 to 2 and 1 to 4 (*Figure 2*f) and weighted Schmid factor ($m_W$) loaded along [100] direction. The full table of common Sn slip systems and weighted Schmid factor (shortlist to $m_W \geq 0.25$) distribution vs. the slip system (#) is presented in supplementary Table S1 and Figure S2.